# BAR PARAMETERS FROM H$\alpha$ OBSERVATIONS

O. Hernandez[1,2], C. Carignan[1], P. Amram[2] and O. Daigle[1,2]
[1]*LAE, Université de Montréal, C.P. 6128, Succ. centre ville, Montréal, Qué. H3C 3J7, Canada*
[2]*OAMP, 2 Place Le Verrier, F–13248 Marseille Cedex 04, France*

**Abstract**   The very first results of the application of the Tremaine-Weinberg (hereafter TW) method on H$\alpha$ velocity fields are presented to find pattern speeds of galaxies. The technique is used for a sample of four barred galaxies and M51. For two of these, our results are very similar to those obtained using the same technique but based on CO observations. For the three other galaxies, a comparison is provided between this observational method and numerical ones.

## 1. Tremaine-Weinberg method with H$\alpha$ velocity fields

The pattern speed $\Omega_p$ of bars or spiral pattern is one of the most important parameter to understand the kinematics and the dynamics of spiral galaxies. On one hand, a dynamical method has been proposed by Garcia-Burillo et al. (1993) to model the gas behaviour in the potential derived from an infrared image of the galaxy for a varying range of $\Omega_p$. On the other hand, three observational methods can be used: the inner resonance 4:1 (Elmegreen et al 1992), the sign inversion of the radial streaming motions across corotation (e.g. the Canzian (1993) test) and the TW (1984) method. The latter is derived from the continuity equation. The tracer (stars, HI, CO, etc...) must follow this equation. For H$\alpha$, we supposed that, to first order, this equation is satisfied for very short times ($\ll$ dynamical time) and we neglect the internal kinematics of HII regions. In that case, in the galactic frame with respect to the galactic center, the luminosity-weighted mean velocity $<\mathcal{V}_{los}>$ in the strip, divided by the component of the luminosity-weighted mean position $<\mathcal{X}>$ parallel to the line of node, is equal to $\Omega_p \sin i$ $(= <\mathcal{V}_{los}>/<\mathcal{X}>)$ .

## 2. Results: pattern speed of five galaxies

Velocity fields and monochromatic images are obtained using 3D adaptive smoothing (Daigle et al, 2004). The position angle (PA), inclination and systemic velocity have been calculated using GIPSY and KARMA (for the detailed description of the method see Hernandez et al. 2004). The figure shows the results obtained for NGC 2903. The fit was performed using a robust Chi-squared model. $\Omega_p$ is found to be 66±7 km/s/kpc. The distance has been chosen such that we can compare with $\Omega_p$ from other sources. The error calculation takes into account the error on the inclination, on the mean line of sight velocity, on the mean position along the line of node and on the PA. The results for the whole sample are shown in the table. The TW method appears to be efficient to derive $\Omega_p$, even if H$\alpha$ does not satisfy the continuity equation for a very long time.

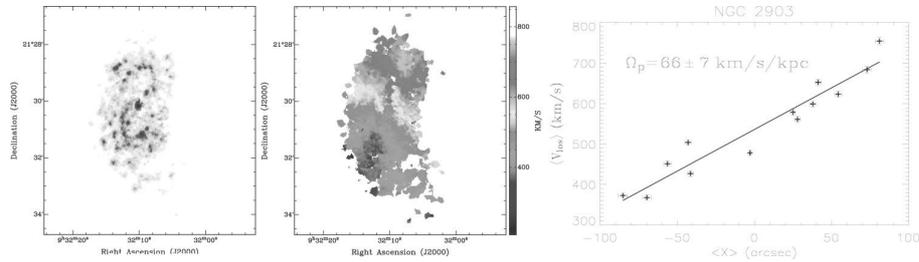

| Galaxy name | Inc (°) | PA (°) | $\Omega_p$ (H$\alpha$) (km/s/arcsec) | $\Omega_p$ (H$\alpha$) (km/s/kpc) | $\Omega_p$ (CO) (km/s/kpc) | $\Omega_p$ (mv)[1] (km/s/kpc) | D (Mpc) |
|---|---|---|---|---|---|---|---|
| NCG 2903 | 61.4 | 20 | 2.3±0.2 | 66±7 | n/a | 68[a] | 7.3 |
| NCG 3359 | 54 | -10 | 2.0±0.3 | 30±4 | n/a | 27[b] | 13.4 |
| NGC 4321 | 31.7 | 153 | 2.6±0.3 | 33±5 | n/a | 20[c] | 16.1 |
| NGC 6946 | 38 | 240 | 1.1±0.2 | 42±6 | 39±9[d] | n/a | 5.5 |
| NGC 5194 | 20 | -10 | 1.8±0.3 | 40±8 | 38±7[d] | 27[e] | 9.5 |

[a] Helfer et al. 2003  [b] Sempere 1999  [c] Sempere et al, 1995  [d] Zimmer et al 2004  [e] Salo & Laurikainen 2003
[1] mv: modeled values